\documentclass[twocolumn,prl,superscriptaddress,showpacs,balancelastpage]{revtex4}
\usepackage[latin1]{inputenc}

\makeatletter

\usepackage{graphicx}
\usepackage{bm}

\makeatother
\begin{document}

\title{Realization of a superconducting atom chip}

\author{T. Nirrengarten}
\author{A. Qarry}
\author{C. Roux}
\author{A. Emmert}
\author{G. Nogues}
\author{M. Brune}
\author{J.-M. Raimond}

\affiliation{Laboratoire Kastler Brossel, D\'{e}partement de Physique de l'Ecole
Normale Sup\'{e}rieure, 24 rue Lhomond, F-75231 Paris Cedex 05, France}

\author{S. Haroche}

\affiliation{Laboratoire Kastler Brossel, D\'{e}partement de Physique de l'Ecole
Normale Sup\'{e}rieure, 24 rue Lhomond, F-75231 Paris Cedex 05, France}

\affiliation{Coll\`{e}ge de France, 11 place Marcelin Berthelot, F-75231 Paris
Cedex 05, France}

\date{\today{}}

\begin{abstract}
We have trapped rubidium atoms in the magnetic field produced by a superconducting atom chip operated at liquid Helium temperatures. Up to $8.2\cdot 10^5$ atoms are held in a Ioffe-Pritchard trap at a distance of 440~$\mu$m from the chip surface, with a temperature of 40~$\mu$K. The trap lifetime reaches 115~s at low atomic densities. These results open the way to the exploration of atom--surface interactions and coherent atomic transport in a superconducting environment, whose properties are radically different from normal metals at room temperature.
\end{abstract}


\maketitle

In atom chip experiments, cold atoms are trapped in the magnetic field gradients created by micron-sized current carrying wires~\cite{MX_SCHMIEDMAYERCHIPREVIEW02} or ferromagnetic structures~\cite{TR_SPREEUWPERMANENTFILMCHIP05}.
The design of the magnetic potential is thus very flexible, allowing for a precise manipulation of the external degrees of freedom of the atomic sample. In this context, the operation of a conveyor belt~\cite{TR_HANSCHCONVEYOR01} or of an atomic beam-splitter~\cite{TR_PRENTISSBEAMSPLITTERCHIP05} have been demonstrated. These devices are particularly interesting for preparing degenerate bosonic or fermionic quantum gases~\cite{MX_HANSCHBECCHIP01, TR_THYWISSENFERMIONCHIP06}.

Atom chips have a wealth of potential applications. They lead to the realization of compact atomic clocks~\cite{TR_REICHELCOHERENCECHIPCLOCK04}, which are a key element for communication and positioning. Integrated atom interferometers~\cite{TR_SCHMIEDMAYERDBLWELL05} could be used as compact inertial sensors. More generally, atom chips open the possibility to bring well-controlled atomic samples close to ``conventional" micro- or optoelectronic systems, an important step for quantum information processing and communication.

These objectives require the preservation of atomic coherence in the vicinity of the chip surface. At room temperature, as is the case in all atom chips experiments so far, various noise sources can jeopardize coherent atomic manipulations. Fluctuating currents of thermal origin (Nyquist noise) in a metallic substrate lead to fluctuating magnetic fields at the trap center. These field fluctuations can induce spurious transitions towards untrapped atomic states and are thus a potentially harmful source of atomic losses~\cite{TR_HENKELMETALNOISE05}.  

The field noise spectrum in the vicinity of a superconductor is expected to be drastically different from that of a normal metal at room temperature. Accordingly, the trap lifetime should increase significantly~\cite{TR_REKDALSUPERCONDNOISE06}, with interesting potentialities for coherent atomic manipulations. Moreover one benefits from an extremely good vacuum because of cryogenic pumping, as it has already been demonstrated for magnetic traps with centimeter-size coils~\cite{TR_LIBBRECHTCRYOTRAP95}.

We report here the operation of a superconducting atom chip. This experiment opens the way to studies of the interactions of cold atoms with superconducting surfaces and currents. In the longer term, we plan to use such a device to prepare cold Rydberg atoms and hold them for long times in a coherence-preserving electric trap~\cite{ENS_RYDBERGTRAP04}. These atoms, shielded from room temperature blackbody radiation, could be employed for quantum information and fundamental studies on atom--surface and atom--atom interactions in dense Rydberg gases~\cite{MX_GALLAGHERCOLD98}.

\begin{figure}[h]
\includegraphics[width=8.5cm]{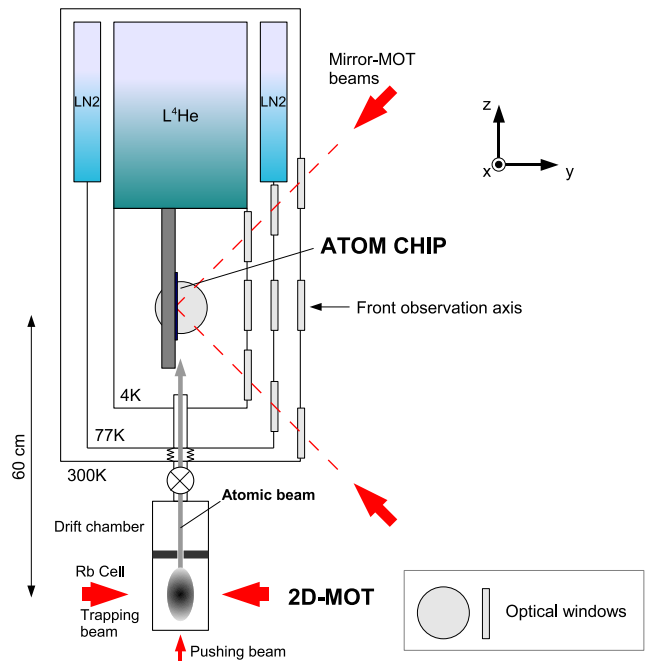}
\caption{\label{fig:schema} (Color online) Scheme of the experiment. Laser beams along the $x$ direction are not represented}
\end{figure}

The atoms are cooled and trapped in a sequential process, taking place in the set-up sketched in Fig.~\ref{fig:schema}. Rubidium atoms ($^{87}$Rb) from a room temperature vapor are trapped and cooled in a 2D-magneto optical trap (MOT) producing an intense slow atomic beam which propagates upwards towards the experimental chamber in the cryostat. The beam is recaptured in front of the chip in a mirror-MOT using centimeter-sized superconducting coils~\cite{MX_SCHMIEDMAYERQUADRUPOLETRAPCHIP04}. The atomic sample is then transferred into a tighter mirror-MOT whose field is produced by on-chip wires and, finally, brought into a Ioffe-Pritchard magnetic trap.

The 2D-MOT UHV chamber at room-temperature is made up of two separate parts (Rb cell and drift chamber) connected together by a 10~mm tunnel with a small 0.8~mm diameter allowing for differential pumping. The pressure in the cell is limited by the Rb background pressure to $\sim 10^{-8}$~mbar whereas the drift chamber is evacuated to a few $10^{-10}$~mbar. Atoms are trapped in the Rb cell along the horizontal $x$ and $y$ directions by counter-propagating laser beams (30~mW per beam, waist 40~mm along $z$ and 20~mm along $x$ and $y$, detuned by $-2.7\ \Gamma$ from the closed $5S_{1/2},\ F=2$ to $5P_{3/2},\ F=3$ transition).  A repumping laser, tuned to the $F=1$ to $F=2$ transition, is superimposed with the trapping beams. The 2D quadrupolar magnetic field gradient in the $xy$ plane is 10~Gauss/cm. A vertical laser beam along $z$ (intensity 6~mW/cm$^2$) extracts the atoms from the 2D-MOT and pushes them towards the cryostat. When they get out of the horizontal repumping beams, a few cm above the 2D-MOT, they are pumped into the $F=1$ dark state and are no longer pushed. The total atomic flux in the drift chamber is $1.5 \cdot 10^7$~atoms/s. The velocity distribution spreads between 10 and 20~m/s and can be tuned by adjusting the power of the pushing beam. The atomic beam divergence (6~mrad) is determined by the diameter of the tunnel connecting the two chambers which is 18~cm above the 2D-MOT. The resulting beam diameter in front of the chip (60~cm above the 2D-MOT) is 3.5~mm, smaller than the distance to the chip surface (5~mm). A direct contamination of the chip by the atomic beam is thus avoided.


The beam from the 2D-MOT enters a $^4$He cryostat through a 10~mm diameter tube made of Nylon and stainless steel bellows with low thermal conductivity. The tube couples into an inner experimental cell cooled down to 4.2~K which has nearly no connection with the insulation vacuum of the cryostat ($8\cdot 10^{-8}$~mbar), in order to benefit from the extremely efficient cryogenic pumping by its cold surfaces. The optical viewports have a diameter of 60~mm and are carefully mounted in order to reduce the stress due to pressure and thermal contractions. Moreover the cold windows are made of SF57 glass, which has a particularly low stress-induced birefringence~\cite{MX_MCMILLANELLIPSOMETER04}. The total measured phase shift between orthogonal polarizations for a laser beam crossing the whole setup is below 0.3~rad (95~\% of the power remains in the desired polarization). 

\begin{figure}
\includegraphics[width=6cm]{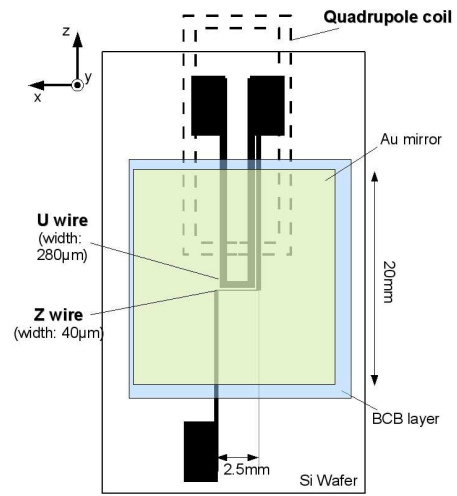}
\caption{\label{fig:chip} (Color online) Layout of the atom chip. Black lines correspond to Nb wires. The dashed lines show the relative position of the ``quadrupole'' coil placed 1.5 mm behind the chip surface.}
\end{figure}

In the inner experimental cell, an ensemble of centimeter-sized superconducting coils mounted around the chip creates a uniform bias field along any direction. Atoms from the low velocity beam are recaptured in a mirror-MOT in front of the chip reflecting surface~\cite{MX_SCHMIEDMAYERQUADRUPOLETRAPCHIP04}. Two counter-propagating laser beams are forming a $45$~deg angle with the chip surface  in the $yz$ plane and two others are propagating in opposite directions along the $x$ axis (see Fig.~\ref{fig:schema}). All beams, brought to the experiment by polarization-maintaining fibers, have a 5~mm waist and a power of 8.5~mW. A fiber-coupled repumping beam (600~$\mu$W) is also sent along the $x$ direction. The quadrupole magnetic field is obtained by an elongated 10~mm $\times$ 28~mm rectangular superconducting coil placed 1.5~mm behind the chip (19 turns of Niomax Nb-Ti wire). It produces near its bottom end the same field as a U-shaped current distribution (see Fig.~\ref{fig:chip}). Combined with a nearly uniform bias field $B_z$ along $z$, it creates a quadrupole field over a spatial extension of a few mm~\cite{MX_SCHMIEDMAYERQUADRUPOLETRAPCHIP04}, well adapted for an efficient capture of the atoms from the slow beam. The current in the rectangular ``quadrupole'' coil, $I_Q$, is 1.77~A. and the bias field, $B_z$, is 3.1~Gauss. 

Observations of the trapped cloud are carried out either by imaging the atomic fluorescence in the MOT beams or by measuring the absorption of a resonant probe beam. In both cases, the image is formed on a cooled CCD camera by lenses located outside the cryostat. The observation direction is either normal to the chip surface (``front" observation, along $y$) or along an axis making an angle of $11$~deg with respect to $x$ (``side" observation). In the latter case, we observe both the direct image of the cloud and its reflection on the chip. We  can thus determine precisely the distance of the cloud from the surface. In a 5~s time interval, we load $5 \cdot 10^7$ atoms into the mirror-MOT at 2.1~mm from the surface. The cloud has a diameter of 2~mm and its temperature, measured by time of flight, is about 300~$\mu$K.

Our atom chip (see Fig.~\ref{fig:chip}) is made on a 65~mm$\times$30~mm silicon wafer (thickness 360~$\mu$m) with a 500~nm insulating oxidized layer. It is coated by a 900~nm thick layer of Nb by cathodic plasma sputtering. A ``U-wire" (width 280~$\mu$m) is used for the on-chip mirror-MOT, and a ``Z-wire" (width 40~$\mu$m) for the magnetic Ioffe-Pritchard trap. The wires and contact pads are produced by standard optical lithography with a soft laser-printed mask followed by reactive ion etching. The resulting  wire edge precision is about 5~$\mu$m. It will be considerably improved in the future using e-beam lithography for critical areas.  Niobium and silicon being bad optical reflectors, the central part of the chip is coated by a 200~nm thick layer of gold (obtained by evaporation) on a 1.5~$\mu$m planarization and insulating layer of BCB (Dow chemicals, Ref. XU35133). The wafer is glued with silver lacquer and mechanically clamped on a copper plate (thickness 1~mm), in thermal contact with the 4.2~K bath. This plate is cut in many places in order to reduce the effect of eddy currents and hence allow for a fast switching of the magnetic fields.  The current is fed in the chip structures through Nb-Ti Niomax wires soldered by a superconducting alloy~\cite{MX_GARFIELDSUPERSOLDERING97} on the contact pads. The critical current for the U-wire circuit is above 5~A. For the Z-wire, the transition to normal state occurs at 1.94(1)~A without laser light. In the presence of the trapping beams, this critical current is reduced to 1.71(1)~A by the local heating due to losses in the gold mirror. 

\begin{figure}
\includegraphics[width=8cm]{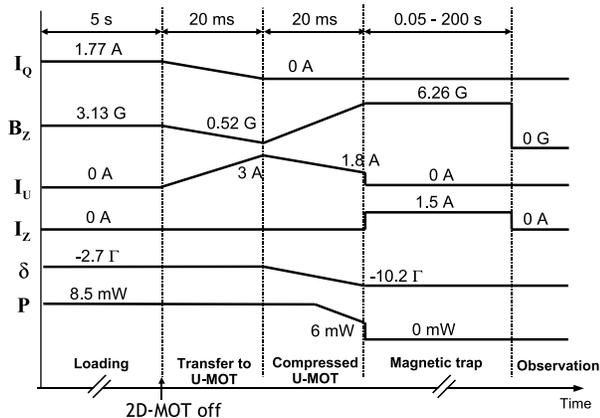}
\caption{\label{fig:sequence} Timing of the experiment. $I_Q$: current in the quadrupole coil, $B_z$: bias field along $z$ direction, $I_U$: current in the U wire, $I_Z$: current in the Z wire, $\delta$: trapping beam detuning, $P$: trapping beam power.}
\end{figure}

The complete timing for a trapping sequence is shown on Fig.~\ref{fig:sequence}. The mirror-MOT loading lasts 5~s. The 2D-MOT lasers are switched off 100~ms before the end of the loading, allowing the slowest atoms to reach the mirror-MOT. The trapped atoms are then transferred in 20~ms into the on-chip mirror MOT, whose magnetic field is produced by the U-wire and the uniform bias (the laser parameters are left unchanged). We decrease the current $I_Q$ linearly from 1.77~A down to 0~A, increase simultaneously the U-wire current $I_U$ up to 3~A, and reduce the bias field $B_z$ down to 0.52~Gauss. These parameters optimize the ``mode matching'' between the two mirror-MOTs. We obtain a transfer efficiency of $\approx85$~\% and the atomic cloud is finally at 1.8~mm from the chip surface.

In the next step, lasting also 20~ms, we bring the sample at a distance of 460~$\mu$m from the chip surface, corresponding to final values of $B_Z$ and $I_U$ of 6.26~Gauss and 1.8~A respectively. As a result, the quadrupole magnetic field gradient increases from 5.8~Gauss/cm to 500~Gauss/cm and the atomic cloud is compressed. In order to reduce atom losses during the compression and to cool down the sample, the red-shift of the trapping beams is increased from $2.7\ \Gamma$ to $10.2\ \Gamma$ and their power $P$ is reduced from 8.5 to 6~mW. At the end of the process, the sample contains $1.2\cdot10^7$ atoms at a temperature of $80$~$\mu$K. Its dimensions are 380~$\mu$m along the $y$ and $z$ directions and 1200~$\mu$m along $x$. We note that this compression stage brings the atoms closer to the chip than typical atom-chip experiments at room temperature. This is because of the limited critical current in the Z-wire which prevents us from performing an efficient transfer from the mirror-MOT to the magnetic trap at atom--chip distances larger than 500~$\mu$m.

\begin{figure}
\includegraphics[width=6.9cm]{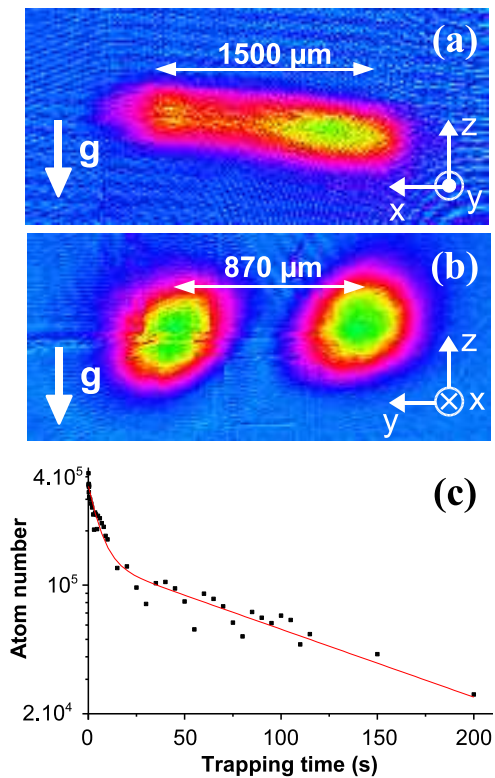}
\caption{\label{fig:trap} (Color online) Absorption images of the atomic cloud after 50~ms in the Ioffe-Pritchard trap with $B_x=0$~Gauss: (a) from the front and (b) from the side (direct image and its reflection in the chip). The $g$ arrows indicate the direction of gravity. The cloud longitudinal and transverse dimensions are 1500~$\mu$m and 156~$\mu$m respectively. The distance from the surface is 430~$\mu$m. (c) Number of atoms in the trap with $B_x=2.75$~Gauss (log scale) as a function of time. The solid line is a bi-exponential fit with characteristic times $\tau_1=5.7(9)$~s and $\tau_2=115(35)$~s.}
\end{figure}

The transfer to the Ioffe-Pritchard trap is finally realized by switching off rapidly all the laser beams in $\sim 1$~$\mu$s. At the same time we cut $I_U$ in $\sim 100$~$\mu$s and set $I_Z$ at a value of 1.5~A (100~$\mu$s rise time) 
The bias field $B_z$ remains constant during this operation. We can additionally switch on an homogeneous bias field along the $+x$ direction. It allows us to control both the value of the magnetic field as well as its gradient in the vicinity of the trap center, two important parameters determining Majorana losses. The magnetic trap is activated for a minimum time of 50~ms in order to let the hottest untrapped atoms escape. All the magnetic fields are then switched off and an absorption image is taken. Figure~\ref{fig:trap}(a) and (b) present front and side absorption images for $B_x=0$~Gauss. The total atom number is $8.2 \cdot 10^5$, with a distance to the chip surface of 440~$\mu$m and a cloud temperature of 40~$\mu$K. The transfer efficiency from the compressed on chip mirror-MOT is thus 7\%. It could be noticeably increased by adding a molasses cooling and optical pumping stage before the final transfer in the magnetic trap. This requires however a fast switching of the bias field $B_z$ which is impossible for the moment because of the large inductance of the bias coils. High-voltage fast-switching power supplies will make it possible.

Figure \ref{fig:trap}(c) shows the number of atoms in the trap as a function of time with $B_x=2.75$~Gauss. It displays a double exponential decay. At the beginning, atom losses are fast (lifetime $\tau_1=5.7(9)$~s). We interpret this as an evaporation process, in which the hottest atoms of the sample escape the trap and stick to the chip surface~\cite{MX_HANSCHCHIP99}. As a confirmation, successive time of flight measurements during the first 10~s show an exponential decrease of the cloud temperature from 40~$\mu$K down to 20~$\mu$K with a characteristic time of $3.2 \pm 1.3$~s. At long times, above 20~s, about 25~\% of the atoms are still trapped. The losses then become dramatically small (lifetime $\tau_2=115(35)$~s). This long lifetime corresponds to an extremely good vacuum in the experimental chamber resulting from the efficient cryogenic pumping by the cold surfaces. By assuming that the vacuum is limited by He residual pressure with a Rb-He scattering cross-section of 100~\AA{}$^2$\cite{TR_DOYLEBUFFERHE02} we infer a pressure of about $3\cdot 10^{-11}$~mbar. This lifetime is large enough to perform evaporative cooling for the production of a BEC~\cite{MX_HANSCHBECCHIP01}. We also observe that a smaller $B_x$ bias results in a larger initial number of atoms (deeper trap) but reduces the value of $\tau_2$, probably because of Majorana losses. For $B_x=0$~Gauss we obtained $\tau_2=5(2)$~s.

In conclusion, our results demonstrate the feasibility of superconducting atom chip traps. We have obtained very long trapping time, in the minute range. The number of trapped atoms is sufficient for reaching Bose-Einstein condensation, a short-term objective of this experiment. A significant increase of the atom number is expected after straightforward improvements. This will allow us to study coherent atomic manipulations in the vicinity of a superconducting environment. In particular we plan to move the cloud away from the initial trapping region towards another part of the chip where the atomic sample will be held in front of a type I or type II superconducting surface. The trapping lifetime or the coherent transport of a BEC sample over these surfaces will be of particular interest.

\begin{acknowledgments}
Laboratoire Kastler Brossel is a laboratory of Universit\'{e} Pierre et Marie Curie and ENS, associated to CNRS (UMR 8552). We acknowledge support of the European Union (CONQUEST and SCALA projects), of the Japan Science and Technology corporation (International Cooperative Research Project~: ``Quantum Entanglement''), and of the R\'egion Ile de France (IFRAF consortium). We had very helpfull discussions with J. Reichel and D. Gu\'ery-Odelin (LKB, ENS). We thank D. Est\`eve and D. Vion (Quantronics group, CEA/SPEC, Saclay) for their extremely valuable help for the chip, for the loan of equipment and for letting us use their clean room facility. M. Rosticher (ENS) greatly helped us for the chip realization.
\end{acknowledgments}



\begin{thebibliography}{24}
\expandafter\ifx\csname natexlab\endcsname\relax\def\natexlab#1{#1}\fi
\expandafter\ifx\csname bibnamefont\endcsname\relax
  \def\bibnamefont#1{#1}\fi
\expandafter\ifx\csname bibfnamefont\endcsname\relax
  \def\bibfnamefont#1{#1}\fi
\expandafter\ifx\csname citenamefont\endcsname\relax
  \def\citenamefont#1{#1}\fi
\expandafter\ifx\csname url\endcsname\relax
  \def\url#1{\texttt{#1}}\fi
\expandafter\ifx\csname urlprefix\endcsname\relax\def\urlprefix{URL }\fi
\providecommand{\bibinfo}[2]{#2}
\providecommand{\eprint}[2][]{\url{#2}}

\bibitem[{\citenamefont{Folman et~al.}(2002)\citenamefont{Folman, Kr{\"u}ger,
  Schmiedmayer, Denschlag, and Henkel}}]{MX_SCHMIEDMAYERCHIPREVIEW02}
\bibinfo{author}{\bibfnamefont{R.}~\bibnamefont{Folman}}~\textit{et~al.},
  \bibinfo{journal}{Advances in Atomic, Molecular and Optical Physics}
  \textbf{\bibinfo{volume}{48}}, \bibinfo{pages}{263} (\bibinfo{year}{2002}).

\bibitem[{\citenamefont{Barb et~al.}(2005)\citenamefont{Barb, Gerritsma, Xing,
  Goedkoop, and Spreeuw}}]{TR_SPREEUWPERMANENTFILMCHIP05}
\bibinfo{author}{\bibfnamefont{E.~A.} \bibnamefont{Hinds}},
 \bibnamefont{and} \bibinfo{author}{\bibfnamefont{I.~G.}
 \bibnamefont{Hughes}}, 
\bibinfo{journal}{J. Phys. D}
  \textbf{\bibinfo{volume}{32}}, \bibinfo{pages}{R119} (\bibinfo{year}{1999});
\bibinfo{author}{\bibfnamefont{I.}~\bibnamefont{Barb}}~\textit{et~al.},
\bibinfo{journal}{Eur. Phys. J. D}
  \textbf{\bibinfo{volume}{35}}, \bibinfo{pages}{75} (\bibinfo{year}{2005});
\bibinfo{author}{\bibfnamefont{A.}~\bibnamefont{Jaakkola}}~\textit{et~al.},
  \bibinfo{journal}{Eur. Phys. J. D} \textbf{\bibinfo{volume}{35}},
  \bibinfo{pages}{81} (\bibinfo{year}{2005});
\bibinfo{author}{\bibfnamefont{C.~D.~J.} \bibnamefont{Sinclair}}~\textit{et~al.},
  \bibinfo{journal}{Phys. Rev. A} \textbf{\bibinfo{volume}{72}},
  \bibinfo{pages}{031603(R)} (\bibinfo{year}{2005}).

\bibitem[{\citenamefont{H{\"a}nsel
  et~al.}(2001{\natexlab{a}})\citenamefont{H{\"a}nsel, Reichel, Hommelhoff, and
  H{\"a}nsch}}]{TR_HANSCHCONVEYOR01}
\bibinfo{author}{\bibfnamefont{W.}~\bibnamefont{H{\"a}nsel}}~\textit{et~al.},
  \bibinfo{journal}{Phys. Rev. Lett.} \textbf{\bibinfo{volume}{86}},
  \bibinfo{pages}{608} (\bibinfo{year}{2001}{\natexlab{a}}).

\bibitem[{\citenamefont{Shin et~al.}(2005)\citenamefont{Shin, Sanner, Jo,
  Pasquini, Saba, Ketterle, Pritchard, Vengalattore, and
  Prentiss}}]{TR_PRENTISSBEAMSPLITTERCHIP05}
\bibinfo{author}{\bibfnamefont{Y.}~\bibnamefont{Shin}}~\textit{et~al.},
  \bibinfo{journal}{Phys. Rev. A} \textbf{\bibinfo{volume}{72}},
  \bibinfo{pages}{021604(R)} (\bibinfo{year}{2005}).

\bibitem[{\citenamefont{H{\"a}nsel
  et~al.}(2001{\natexlab{b}})\citenamefont{H{\"a}nsel, Hommelhoff, H{\"a}nsch,
  and Reichel}}]{MX_HANSCHBECCHIP01}
\bibinfo{author}{\bibfnamefont{W.}~\bibnamefont{H{\"a}nsel}}~\textit{et~al.},
  \bibinfo{journal}{Nature (London)} \textbf{\bibinfo{volume}{413}},
  \bibinfo{pages}{498} (\bibinfo{year}{2001}{\natexlab{b}});
\bibinfo{author}{\bibfnamefont{H.}~\bibnamefont{Ott}}~\textit{et~al.},
  \bibinfo{journal}{Phys. Rev. Lett.} \textbf{\bibinfo{volume}{87}},
  \bibinfo{pages}{230401} (\bibinfo{year}{2001}).

\bibitem[{\citenamefont{Aubin et~al.}(2006)\citenamefont{Aubin, Myrskog,
  Extavour, LeBlanc, McKay, Stummer, and
  Thywissen}}]{TR_THYWISSENFERMIONCHIP06}
\bibinfo{author}{\bibfnamefont{S.}~\bibnamefont{Aubin}}~\textit{et~al.},
  \bibinfo{journal}{Nature Physics} \textbf{\bibinfo{volume}{2}},
  \bibinfo{pages}{384} (\bibinfo{year}{2006}).

\bibitem[{\citenamefont{Treutlein et~al.}(2004)\citenamefont{Treutlein,
  Hommelhoff, Steinmetz, H{\"a}nsch, and
  Reichel}}]{TR_REICHELCOHERENCECHIPCLOCK04}
\bibinfo{author}{\bibfnamefont{P.}~\bibnamefont{Treutlein}}~\textit{et~al.},
  \bibinfo{journal}{Phys. Rev. Lett.} \textbf{\bibinfo{volume}{92}},
  \bibinfo{pages}{203005} (\bibinfo{year}{2004});
\bibinfo{author}{\bibfnamefont{S.}~\bibnamefont{Knappe}}~\textit{et~al.},
  \bibinfo{journal}{Optics Express} \textbf{\bibinfo{volume}{13}},
  \bibinfo{pages}{1249} (\bibinfo{year}{2005}).

\bibitem[{\citenamefont{Schumm et~al.}(2005)\citenamefont{Schumm, Hofferberth,
  Andersson, Wildermuth, Groth, Bar-Joseph, Schmiedmayer, and
  Kruger}}]{TR_SCHMIEDMAYERDBLWELL05}
\bibinfo{author}{\bibfnamefont{T.}~\bibnamefont{Schumm}}~\textit{et~al.},
  \bibinfo{journal}{Nature Physics} \textbf{\bibinfo{volume}{1}},
  \bibinfo{pages}{57} (\bibinfo{year}{2005});
\bibinfo{author}{\bibfnamefont{Y.}~\bibnamefont{Wang}}~\textit{et~al.},
\bibinfo{journal}{Phys. Rev. Lett.} \textbf{\bibinfo{volume}{94}}, \bibinfo{pages}{90405}
  (\bibinfo{year}{2005}).


\bibitem[{\citenamefont{Henkel}(2005)}]{TR_HENKELMETALNOISE05}
\bibinfo{author}{\bibfnamefont{C.}~\bibnamefont{Henkel}},
  \bibinfo{journal}{Eur. Phys. J. D} \textbf{\bibinfo{volume}{35}},
  \bibinfo{pages}{59} (\bibinfo{year}{2005}).

\bibitem[{\citenamefont{Skagerstam et~al.}(2006)\citenamefont{Skagerstam,
  Hohenester, Eiguren, and Rekdal}}]{TR_REKDALSUPERCONDNOISE06}
\bibinfo{author}{\bibfnamefont{S.}~\bibnamefont{Scheel}}~\textit{et~al.},
  \bibinfo{journal}{Phys. Rev. A} \textbf{\bibinfo{volume}{72}},
  \bibinfo{pages}{042901} (\bibinfo{year}{2005});
\bibinfo{author}{\bibfnamefont{B.}~\bibnamefont{Ska\-ger\-stam}}~\textit{et~al.},
  \bibinfo{journal}{Phys. Rev. Lett.} \textbf{\bibinfo{volume}{97}},
  \bibinfo{pages}{070401} (\bibinfo{year}{2006}).

\bibitem[{\citenamefont{Willems and Libbrecht}(1995)}]{TR_LIBBRECHTCRYOTRAP95}
\bibinfo{author}{\bibfnamefont{P.~A.} \bibnamefont{Willems}} \bibnamefont{and}
  \bibinfo{author}{\bibfnamefont{K.~G.} \bibnamefont{Libbrecht}},
  \bibinfo{journal}{Phys. Rev. A} \textbf{\bibinfo{volume}{51}},
  \bibinfo{pages}{1403} (\bibinfo{year}{1995}).

\bibitem[{\citenamefont{Hyafil et~al.}(2004)\citenamefont{Hyafil, Mozley,
  Perrin, Tailleur, Nogues, Brune, Raimond, and Haroche}}]{ENS_RYDBERGTRAP04}
\bibinfo{author}{\bibfnamefont{P.}~\bibnamefont{Hyafil}}~\textit{et~al.},
  \bibinfo{journal}{Phys. Rev. Lett.} \textbf{\bibinfo{volume}{93}},
  \bibinfo{pages}{103001} (\bibinfo{year}{2004}).

\bibitem[{\citenamefont{Anderson et~al.}(1998)\citenamefont{Anderson, Veale,
  and Gallagher}}]{MX_GALLAGHERCOLD98}
\bibinfo{author}{\bibfnamefont{W.~R.} \bibnamefont{Anderson}},
   \bibinfo{author}{\bibfnamefont{J.~R.} \bibnamefont{Veale}}, \bibnamefont{and}
   \bibinfo{author}{\bibfnamefont{T.~F.} \bibnamefont{Gallagher}},
  \bibinfo{journal}{Phys. Rev. Lett.} \textbf{\bibinfo{volume}{80}},
  \bibinfo{pages}{249} (\bibinfo{year}{1998});
\bibinfo{author}{\bibfnamefont{T.~J.} \bibnamefont{Carroll}}~\textit{et~al.},
  \bibinfo{journal}{Phys. Rev. Lett.} \textbf{\bibinfo{volume}{93}},
  \bibinfo{pages}{153001} (\bibinfo{year}{2004}).

\bibitem[{\citenamefont{Wildermuth et~al.}(2004)\citenamefont{Wildermuth,
  Kr{\"u}ger, Becker, Brajdic, Haupt, Kasper, Folman, and
  Schmiedmayer}}]{MX_SCHMIEDMAYERQUADRUPOLETRAPCHIP04}
\bibinfo{author}{\bibfnamefont{S.}~\bibnamefont{Wildermuth}}~\textit{et~al.},
  \bibinfo{journal}{Phys. Rev. A} \textbf{\bibinfo{volume}{69}},
  \bibinfo{pages}{030901(R)} (\bibinfo{year}{2004}).

\bibitem[{\citenamefont{McMillan et~al.}(2004)\citenamefont{McMillan, Taborek,
  and Rutledge}}]{MX_MCMILLANELLIPSOMETER04}
\bibinfo{author}{\bibfnamefont{T.}~\bibnamefont{McMillan}},
  \bibinfo{author}{\bibfnamefont{P.}~\bibnamefont{Taborek}}, \bibnamefont{and}
  \bibinfo{author}{\bibfnamefont{J.~E.} \bibnamefont{Rutledge}},
  \bibinfo{journal}{Rev. Sci. Inst.} \textbf{\bibinfo{volume}{75}},
  \bibinfo{pages}{5005} (\bibinfo{year}{2004}).

\bibitem[{\citenamefont{Garfield}(1997)}]{MX_GARFIELDSUPERSOLDERING97}
\bibinfo{author}{\bibfnamefont{J.~R.~B.} \bibnamefont{Garfield}},
  \bibinfo{journal}{Rev. Sci. Inst.} \textbf{\bibinfo{volume}{68}},
  \bibinfo{pages}{1906} (\bibinfo{year}{1997}).

\bibitem[{\citenamefont{Reichel et~al.}(1999)\citenamefont{Reichel, H{\"a}nsel,
  and H{\"a}nsch}}]{MX_HANSCHCHIP99}
\bibinfo{author}{\bibfnamefont{J.}~\bibnamefont{Reichel}},
  \bibinfo{author}{\bibfnamefont{W.}~\bibnamefont{H{\"a}nsel}},
  \bibnamefont{and} \bibinfo{author}{\bibfnamefont{T.~W.}
  \bibnamefont{H{\"a}nsch}}, \bibinfo{journal}{Phys. Rev. Lett.}
  \textbf{\bibinfo{volume}{83}}, \bibinfo{pages}{3398} (\bibinfo{year}{1999}).

\bibitem[{\citenamefont{Egorov et~al.}(2002)\citenamefont{Egorov, Lahaye,
  Sch{\"o}llkopf, Friedrich, and Doyle}}]{TR_DOYLEBUFFERHE02}
\bibinfo{author}{\bibfnamefont{D.}~\bibnamefont{Egorov}}~\textit{et~al.},
  \bibinfo{journal}{Phys. Rev. A} \textbf{\bibinfo{volume}{66}},
  \bibinfo{pages}{043401} (\bibinfo{year}{2002}).

\end{thebibliography}

\end{document}